\documentclass[aps,prl,floats,floatfix,twocolumn,showpacs,superscriptaddress]{revtex4}
\usepackage{latexsym,amsmath}
\usepackage{amsbsy}
\usepackage[pdftex]{graphicx}
\usepackage{amssymb}
\usepackage{epstopdf}
\usepackage[usenames]{color}

\begin{document}

\title{Epidemic spreading on interconnected networks}

\author{Anna Saumell-Mendiola}
\affiliation{Departament de F{\'\i}sica Fonamental, Universitat de Barcelona, Mart\'{\i} i Franqu\`es 1, 08028 Barcelona, Spain}

\author{M. \'Angeles Serrano}

\affiliation{Departament de Qu\'imica F\'isica, Universitat de Barcelona, Mart\'i i Franqu\`es 1, 08028, Barcelona, Spain}

\author{Mari{\'a}n Bogu{\~n}{\'a}}

\affiliation{Departament de F{\'\i}sica Fonamental, Universitat de Barcelona, Mart\'{\i} i Franqu\`es 1, 08028 Barcelona, Spain}

\date{\today}

\begin{abstract}

Many real networks are not isolated from each other but form networks of networks, often interrelated in non trivial ways. Here, we analyze an epidemic spreading process taking place on top of two interconnected complex networks. We develop a heterogeneous mean field approach that allows us to calculate the conditions for the emergence of an endemic state. Interestingly, a global endemic state may arise in the coupled system even though the epidemics is not able to propagate on each network separately, and even when the number of coupling connections is small. Our analytic results are successfully confronted against large-scale numerical simulations.
\end{abstract}


\maketitle

Epidemic spreading is one of the most successful application areas of the new science of networks~\cite{Barrat:2008bh,NewmanBook:2010}. Indeed, the general acceptance within the scientific community that many diseases, like sexually transmitted diseases or the H1N1 virus, spread over networked systems represents a major step toward their understanding and control~\cite{Lloyd:2001vh,Liljeros:2001fy,Balcan:2009}. From a physics perspective, epidemic processes have been widely studied as a paradigm of non-equilibrium phase transitions with absorbing states~\cite{Marro:1999bw}. When applied to complex networks, these processes have become a source of new and striking phenomena that do not have a counterpart in regular lattices. Germane examples are the absence of epidemic and percolation thresholds in scale-free networks with a power law degree distribution $P(k) \sim k^{-\gamma}$ with $\gamma\in (2,3]$, and an anomalous critical behavior when $\gamma \in (3,4)$~\cite{Pastor-Satorras:2001oi,Pastor-Satorras:2001tt,Cohen:2002jx}.

We currently have a solid understanding of epidemic processes when they take place on single isolated networks. In contrast, our comprehension is very limited when epidemics happen on coupled interconnected networks. For example, sexually transmitted diseases can propagate both in heterosexual and homosexual networks of sexual contacts~\citep{Liljeros:2001fy}. These two networks are not completely isolated due to the existence of bisexual individuals, which act as an effective coupling between the two networks and potentially affect their epidemic properties~\cite{Jeffries:2011}. To the best of our knowledge, a theory describing these type of systems has not yet been fully developed.

In this paper, we fill this gap and present a rigorous heterogeneous mean field study of the susceptible-infected-susceptible (SIS) model taking place on two interconnected complex networks. Our analysis reveals a highly non-trivial behavior of the epidemic process depending on the strength and nature of the coupling between the networks. We calculate the global epidemic threshold of the process, which turns out to be smaller than the epidemic thresholds of the two networks separately under certain conditions. This implies that {\it an endemic state may arise even if the epidemics is not endemic in any of the two networks separately}, as we prove analytically and with large-scale computer simulations.

To begin our analysis, we have to specify the topological properties of our networks. Let A and B be two interconnected random networks with given local properties and given two-points correlations. Each node in network A (respectively B) is characterized by a vector degree $\mathbf{k}_a\equiv(k_{aa},k_{ab})$ representing the number of its internal connections $k_{aa}$ to other nodes in A, and the number of external connections $k_{ab}$ with nodes of network B. The analogous to the degree distribution in single networks is then the probability that a randomly chosen node of network A has a vector degree $\mathbf{k}_a$, $P_A(\mathbf{k}_a)=P_A(k_{aa},k_{ab})$, where, in general, $k_{aa}$ and $k_{ab}$ may be correlated. A similar definition applies to $P_B(\mathbf{k}_b)$ for network B. These two distributions are arbitrary except for the consistency condition $N_A/N_B= \langle k_{ba} \rangle/\langle k_{ab} \rangle$, where $N_A/N_B$ is the ratio between the sizes of the two networks and the right hand side is the ratio between their external average degrees. This condition states that the total number of edges leaving network A towards B must be the same going from network B to network A. Two-points correlations, on the other hand, are encoded by the transition probabilities $P_{AA}(\mathbf{k}'_a|\mathbf{k}_a)$, $P_{AB}(\mathbf{k}'_b|\mathbf{k}_a)$, $P_{BA}(\mathbf{k}'_a|\mathbf{k}_b)$,and $P_{BB}(\mathbf{k}'_b|\mathbf{k}_b)$. For instance, $P_{AB}(\mathbf{k}'_b|\mathbf{k}_a)$ is the probability that, being in a node of network A with vector degree $\mathbf{k}_a$ a randomly chosen neighbor in network B has vector degree $\mathbf{k}'_b$. Similar definitions applies to the rest of the transition probabilities.

As the epidemic spreading model, we consider the SIS model that, together with the susceptible-infected-recovered model (SIR), is one of the best studied models in epidemiology~\cite{May:1991qo}. The model has a non-equilibrium phase transition between an endemic state with sustained epidemic activity and a healthy phase where the epidemics dies out. Individuals can be in two different states, either susceptible ($S$) or infected ($I$). Infected individuals decay spontaneously to the susceptible state at rate $\delta$ (that, without loss of generality, we set to $\delta=1$) whereas susceptible ones get infected at a rate proportional to the number of infected neighbors they have at a given time. For two interconnected networks, we have to specify these processes separately. Let $\lambda^{aa}$ ($\lambda^{bb}$) be the infectious rate between nodes in network A (B) and $\lambda^{ab}$ ($\lambda^{ba}$) the infectious rate from a node in A (B) to a node in B (A).

The quantity of interest in the SIS dynamics is the prevalence, $\rho(t)$, defined as the fraction of infected nodes at a given time. To describe its evolution, we use the heterogeneous mean field approximation. In this approximation, nodes are classified within classes of equivalence such that all nodes within a given class are considered as statistically equivalent. In our case, classes of equivalence are defined by the network itself (A or B) and by specific values of the vector degree. Following these assumptions, we define the partial prevalences $\rho^A_{\mathbf{k}_a}(t)$ and $\rho^B_{\mathbf{k}_b}$ as the fraction of infected nodes with a given vector degree. The total prevalences for each network is then computed as $\rho^A(t)=\sum_{\mathbf{k}_a}P_A(\mathbf{k}_a) \rho^A_{\mathbf{k}_a}(t)$ and $\rho^B(t)=\sum_{\mathbf{k}_b}P_B(\mathbf{k}_b) \rho^B_{\mathbf{k}_b}$. Following~\cite{Pastor-Satorras:2001oi,Boguna:2002zl}, the time evolution of the dynamics can be written as
\begin{equation}
\begin{split}
\frac{d \rho^A_{\mathbf{k}_a}}{dt}= & - \rho^A_{\mathbf{k}_a} + \lambda^{aa} (1- \rho^A_{\mathbf{k}_a}) k_{aa}  \sum _{\mathbf{k}'_a} \rho^A_{\mathbf{k}'_a} P_{AA} (\mathbf{k}'_a | \mathbf{k}_a )  \\
& +\lambda^{ba}(1- \rho^A_{\mathbf{k}_a})  k_{ab} \sum _{\mathbf{k}'_b} \rho^B_{\mathbf{k}'_b} P_{AB} (\mathbf{k}'_b | \mathbf{k}_a ) \;.
\end{split}
\label{eq:1}
\end{equation}
The first line in Eq.~(\ref{eq:1}) describes the standard SIS model for a single network~\cite{Boguna:2002zl}, whereas the second line appears due to the coupling of network A with network B. An analogous equation can be written for network B by swapping the indices $A \rightarrow B$ and $a \rightarrow b$ in Eq.~(\ref{eq:1}).

As in a single network, this process undergoes a phase transition between a healthy phase with $\rho^A_{\mathbf{k}_a}=\rho^B_{\mathbf{k}_b}=0$ and an endemic phase with  $\rho^A_{\mathbf{k}_a} \neq 0$ and $\rho^B_{\mathbf{k}_b} \neq 0$. However, a mixed phase with endemic activity in one network whereas the other is in a heathy state is not possible in the system formed by the two coupled networks, where the epidemics propagates to the whole system if it is able to propagate in one of the networks. This is due to the fact that the state $\rho^A_{\mathbf{k}_a} = 0$ and $\rho^B_{\mathbf{k}_b} \neq 0$ is not a fixed point of the dynamics in Eq.~(\ref{eq:1}). The critical point separating the healthy and endemic phases can be obtained by studying the stability of the absorbing solution. This can be done by linearizing the system of Eqs.~(\ref{eq:1}) around $\rho^A_{\mathbf{k}_a}=\rho^B_{\mathbf{k}_b}=0$ and studying the spectrum of the corresponding matrix. Close to the absorbing state, Eqs.(\ref{eq:1}) can be written as
\begin{equation}
\frac{d \vec{\rho} }{dt}=-\vec{\rho}+\mathbb{C} \vec{\rho},
\end{equation}
where
we define the vector prevalence as $\vec{\rho}\equiv (\rho^A_{\mathbf{k}_a},\rho^B_{\mathbf{k}_b})$ and
\begin{equation}
\mathbb{C}=
\left(
\begin{array}{cc}
\lambda^{aa} k_{aa} P_{AA} (\mathbf{k}'_a | \mathbf{k}_a ) & 
\lambda^{ba} k_{ab} P_{AB} (\mathbf{k}'_b | \mathbf{k}_a ) \\[0.5cm]
\lambda^{ab} k_{ba} P_{BA} (\mathbf{k}'_a | \mathbf{k}_b ) &
\lambda^{bb} k_{bb} P_{BB} (\mathbf{k}'_b | \mathbf{k}_b )
\end{array}
\right).
\end{equation}
The absorbing state is stable whenever the maximum eigenvalue of matrix $\mathbb{C}$ satisfies $\Lambda_m<1$. Otherwise, the absorbing state is unstable and an endemic state takes over the system. The critical epidemic point is then defined by $\Lambda_m=1$.

When the networks do not have two-points correlations, the transition probabilities can be written as
\begin{equation}
P_{AA} (\mathbf{k}'_a | \mathbf{k}_a ) = \frac{k'_{aa} P(\mathbf{k}'_a)}{\langle k_{aa} \rangle } \text{;  }
P_{AB} (\mathbf{k}'_b | \mathbf{k}_a ) = \frac{k'_{ba} P(\mathbf{k}'_b)}{\langle k_{ba} \rangle }
\end{equation} 
In this case, the eigenvalues of matrix $\mathbb{C}$ are the solutions of the following equation
\begin{equation}\label{eigenvalues}
\begin{split}
[ x -  \Lambda_{A}][ x - \Lambda_{B}][ x^2 - \Lambda^2_{AB}]=\\[0.3cm]
 \alpha_{ab} \alpha_{ba}  + \mu_{ab} \alpha_{ba}[ x - {\Lambda }_{A}] +  \mu_{ba} \alpha_{ab} [x - {\Lambda }_{B}],
\end{split}
\end{equation}
where
\begin{equation}
\Lambda_A=\lambda^{aa}\frac{\langle {k_{aa}}^2 \rangle }{\langle k_{aa} \rangle} \text{, }
\Lambda_B=\lambda^{bb}\frac{\langle {k_{bb}}^2 \rangle }{\langle k_{bb} \rangle} \text{, and }
\label{eigenvaluesAandB}
\end{equation}
\begin{equation}
\Lambda^2_{AB}=\lambda^{ab} \frac{\langle k_{ab}^2 \rangle}{\langle k_{ab} \rangle} \lambda^{ba} \frac{ \langle k_{ba}^2 \rangle}{ \langle k_{ba} \rangle}\equiv \mu_{ab} \mu_{ba},
\end{equation}
and where
\begin{equation}
\alpha_{ab}=\lambda^{aa} \lambda^{ab} \frac{\langle k_{aa} k_{ab} \rangle^2  }{\langle k_{aa} \rangle \langle k_{ab} \rangle},
\alpha_{ba}=\lambda^{bb} \lambda^{ba} \frac{\langle k_{bb} k_{ba} \rangle^2  }{\langle k_{bb} \rangle \langle k_{ba} \rangle}.
\end{equation}
Constants appearing on the left hand side of Eq.~(\ref{eigenvalues}) have a clear interpretation. Indeed, $\Lambda_A$ and $\Lambda_B$ are the maximum eigenvalues of networks A and B as if they were isolated. Therefore if, for instance, $\Lambda_A>1$ then network A is able to sustain an endemic state by itself when isolated from network B. Similarly, $\Lambda_{AB}$ is the maximum eigenvalue of the network AB as a pure bipartite system, that is, when all internal connections inside networks A and B are absent. Again, when $\Lambda_{AB}>1$, the pure bipartite network AB is able to sustain an endemic state, both in the coupled system and even if there were no connections whatsoever within each individual network. Constants $\alpha_{ab}$, $\alpha_{ba}$, $\mu_{ab}$ and $\mu_{ba}$ appearing on the right hand side of Eq.~(\ref{eigenvalues}) contain information about the strength and nature of the coupling between the nets.

From Eq.~(\ref{eigenvalues}), it is easy to see that the maximum eigenvalue of matrix $\mathbb{C}$, $\Lambda_m$, is always larger than $\max{(\Lambda_A,\Lambda_B,\Lambda_{AB})}$. It is therefore possible to find endemic states with   $\Lambda_m>1$ but where $\Lambda_A<1$, $\Lambda_B<1$, and $\Lambda_{AB}<1$, that is, situations where neither networks A and B isolated nor the pure bipartite network AB are able to sustain the endemic state and yet the epidemics pervades in the coupled system. This effect is more or less important depending on the strength of the coupling, {\it i. e.}, the number of links between the two networks --quantified by $\langle k_{ab} \rangle$ and $\langle k_{ba} \rangle$-- and the specific correlations between internal and external degrees, measured by the factors $\langle k_{aa} k_{ab} \rangle$ and $\langle k_{ba} k_{bb} \rangle$. 

In the rest of the paper, we focus on the interesting case $\Lambda_A<1$, $\Lambda_B<1$, and $\Lambda_{AB}<1$, and ask under which condition the endemic state exists, that is , $\Lambda_m>1$. From Eq.~(\ref{eigenvalues}), we see that this happens when the right hand side of Eq.~(\ref{eigenvalues}) evaluated at $x=1$ is larger than the left hand side evaluated at the same point, which after some algebra yields
\begin{equation}
[\alpha_{ab}+\mu_{ab}(1-\Lambda_A)][\alpha_{ba}+\mu_{ba}(1-\Lambda_B)]>(1-\Lambda_A)(1-\Lambda_B).
\label{critical_line}
\end{equation}
This equation is one of the main results of our paper. It allows us to evaluate the conditions for the emergence of the endemic state in many different situations. 

In real networks of sexual contacts, the most promiscuous individuals in one network are also the ones with the largest number of sexual partners in the other network~\citep{Jeffries:2011}. This fact suggests that there exists a positive correlation between the internal and external degree for a given node. We model these correlations by taking the vector degree distribution to be $P_A(k_{aa},k_{ab})=P_A(k_{aa})g(k_{ab}|k_{aa})$, where $g(k_{ab}|k_{aa})$ is a Poisson distribution with mean $\bar{k}_{ab}(k_{aa})=\langle k_{ab} \rangle k_{aa}^{\beta}/\langle k_{aa}^{\beta} \rangle$, and analogously for $P_B(k_{bb},k_{ba})$. This choice allows us to interpolate between a random distribution of links between the two networks, $\beta=0$, and positive correlations, where high degree nodes in both networks concentrate the majority of the coupling links, $\beta>0$. Inserting this assumption in Eq.~(\ref{critical_line}), we obtain the critical lines that define the phase diagram in the hyperplane $\Lambda_A$, $\Lambda_B$, $\alpha_{ab}$, and $\alpha_{ba}$.
\begin{figure}[h]
\hspace*{-0.5cm}
\centerline{\includegraphics[width=3.4in]{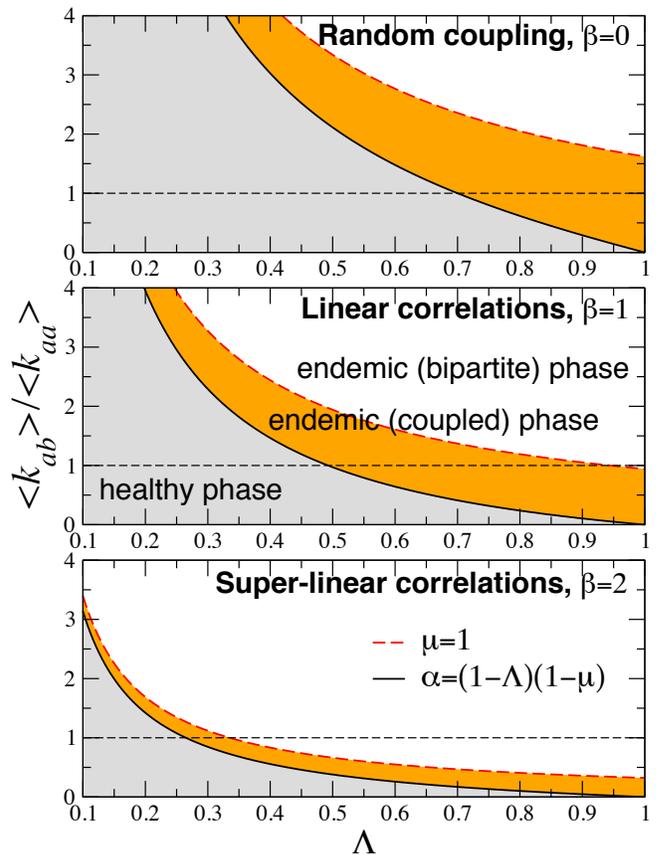}} 
\caption{Phase diagram showing the healthy phase and the endemic phase for the case of a random coupling (top), linear correlations (middle) and super-linear correlations between external and internal degrees (bottom). In all cases, networks A and B are identical, with an exponential degree distribution with $\langle k_{aa} \rangle=\langle k_{bb} \rangle=10$ and minimum degree $2$. The area depicted in gray corresponds to the healthy phase and the transition to the endemic state is represented by the black curve. The orange area corresponds to the endemic state that needs the contribution of both internal and external links and the white area to the case when the pure bipartite network alone is able to sustain the endemic state. The horizontal dashed line marks the point where external connections outnumber internal ones.}
\label{phase_diagram}
\end{figure}

Figure~\ref{phase_diagram} shows examples of phase diagrams for two identical networks with a symmetric coupling, that is, $\Lambda_A=\Lambda_B=\Lambda$, $\alpha_{ab}=\alpha_{ba}=\alpha$, and $\mu_{ab}=\mu_{ba}=\mu$. In this case, Eq.~(\ref{critical_line}) simplifies to $\alpha>(1-\Lambda) (1-\mu)$. We consider the case of networks with an exponential degree distribution and the values of $\beta=0,1,2$. The most interesting case corresponds to the range of parameters where both internal and external links are needed for the existence of the endemic state --depicted in orange in Fig. \ref{phase_diagram}, and particularly when the internal connections outnumber the external ones (the portion below the dashed line). This area shrinks when increasing $\beta$ but, simultaneously, it also appears at lower values of the ratio of external versus internal connections for the same $\Lambda$. Therefore, keeping the two networks unchanged and for a fixed number of links between them, the epidemic can either be in the healthy or endemic phase depending on how these links are distributed among the nodes of the networks. Notice that, if correlations are strong enough, the bipartite network alone is able to sustain by itself the endemic state in the coupled system.
\begin{figure}[h]
\centerline{\includegraphics[width=3.4in]{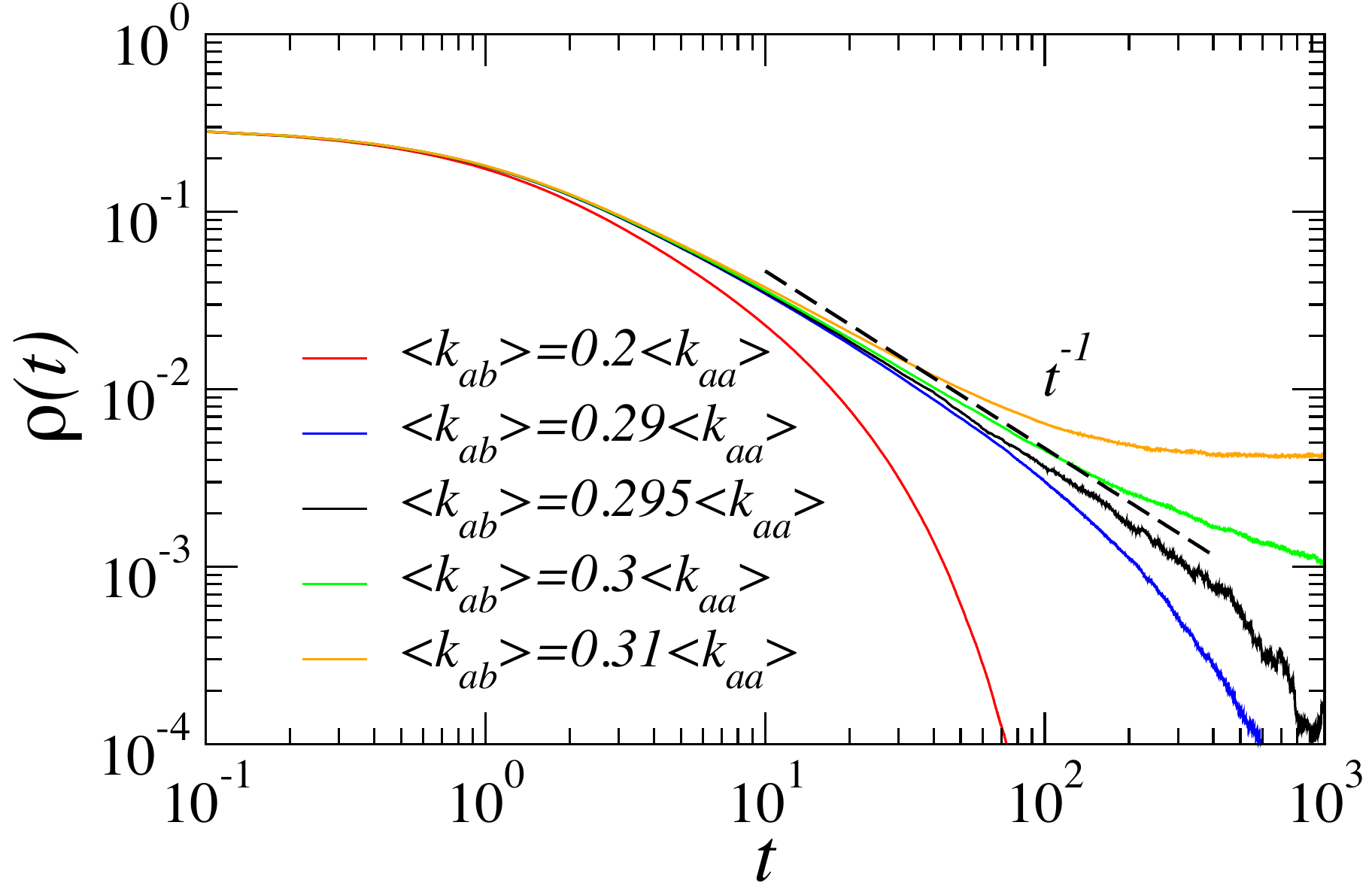}}
\caption{Time evolution of the prevalence $\rho(t)$ below and above the critical point for $\Lambda=0.8$ for two identical networks interconnected with linear correlations, {\it i.~e.}, $\beta=1$. Each network has an exponential degree distribution with $\langle k \rangle=10$ and minimum degree $2$. The size of each network is $N=10^6$ and results are averaged over $100$ different realizations of the process. Right at the critical point, the prevalence decays as $\rho(t)\sim t^{-1}$.}
\label{fig:evolucio}
\end{figure}

We checked our predictions with large scale numerical simulations. The SIS dynamics is simulated with a continuous time dynamics as follows. During the course of the simulation, we keep track of the number of infected nodes $N_I(t)$ and the number of active links $E_A(t)$, where an active link is defined as a link connecting a susceptible and an infected node. At each step, with probability $p_r=N_I(t)(N_I(t)+\lambda E_A(t))^{-1}$, a randomly chosen infected node is turned susceptible whereas, with probability $1-p_r$, an active link is chosen at random and the susceptible node attached to it is turned infected. After this procedure, time is updated as $t \rightarrow t+(N_I(t)+\lambda E_A(t))^{-1}$. We run this algorithm on the networks used in Fig.~\ref{phase_diagram} of size $N=10^6$. Having fixed the internal network properties, we fix the value of $\Lambda$ by adjusting the infectious rate $\lambda$ using Eq.~(\ref{eigenvaluesAandB}). For each value of $\Lambda$, we change the average external degree and study the temporal behavior of the prevalence. The critical point is determined as the point where $\rho(t)$ decays as a power law~\cite{Marro:1999bw}, as shown in Fig.~\ref{fig:evolucio}.  
\begin{figure}[t]
\centerline{\includegraphics[width=3.4in]{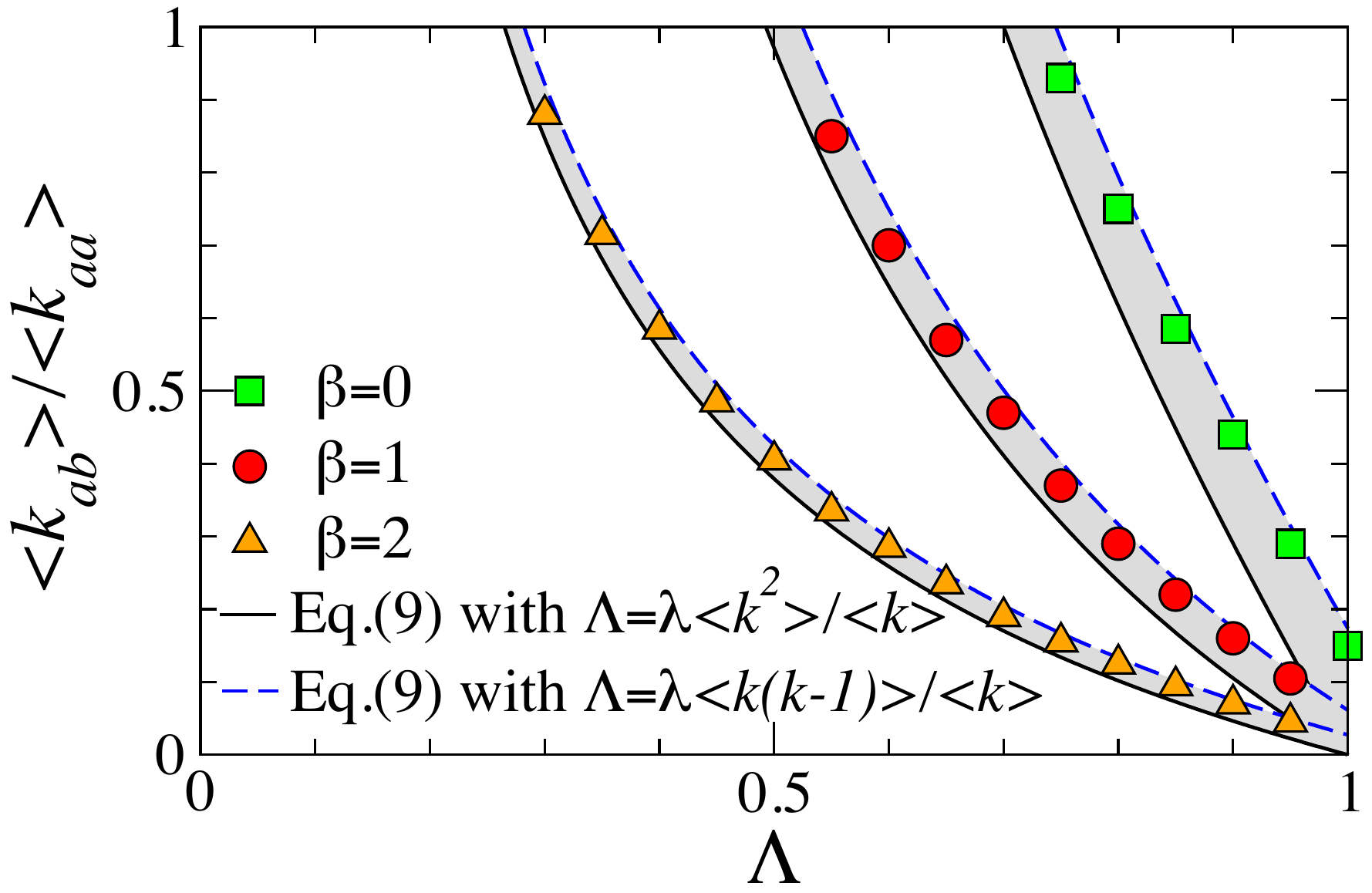}}
\caption{Critical lines for $\beta=0,1,2$ for the same networks as in Fig.~\ref{fig:evolucio}. Dotted lines correspond to numerical simulations, black lines are the theoretical predictions according to Eq.~(\ref{critical_line}), whereas blue dashed lines are obtained by rescaling the $\Lambda$ axis as $\Lambda \rightarrow \Lambda(1-\langle k \rangle/\langle k^2 \rangle)^{-1}$. }
\label{fig:valors_alpha}
\end{figure}

Using this method, we compute the critical line in the plane $(\langle k_{ab} \rangle/\langle k_{aa} \rangle,\Lambda)$ for the three cases analyzed in Fig.~\ref{phase_diagram}. We limit our search to the domain $(\langle k_{ab} \rangle<\langle k_{aa} \rangle,\Lambda<1)$. Simulation results are shown in Fig.~\ref{fig:valors_alpha} along with the analytic predictions given by Eq.~(\ref{critical_line}). We observe a systematic shift between theory and simulations. The reason is that our mean field approach does not consider dynamical correlations. These correlations imply that, with high probability, an infected node has its infecting neighbor still infected during some time right after the infectious event. This reduces the number of potential new infections that this node can produce. We can correct this effect in the same way as it is done in the SIR model, just by reducing the number of contacts by $1$. In our equations, this is achieved by replacing the maximum eigenvalue of the networks by $\Lambda=\lambda  \langle k(k-1) \rangle/\langle k \rangle$. This is, of course, a limiting case because there are cases where the infecting neighbor recovers and can be reinfected. Thus, we expect to find simulations results between these two extremes. We show this correction in Fig.~\ref{critical_line} as the gray area. Indeed, all simulation points fall within this gray area, confirming then our intuition.

The study of interconnected and/or interdependent networks reveals new and unexpected phenomena~\cite{BPPSH10}. Here, we have shown that two networks well below their respective epidemic thresholds may sustain an endemic state when coupling connections are added, even in small number. This may have important implications for the design of efficient control strategies. However, the effects of the coupling are highly non-trivial and may vary depending on the strength and the correlations of the interconnecting links. We foresee similar effects appearing in many different dynamics showing equilibrium and non-equilibrium phase transitions. 

\section*{Acknowledgements}

This work was supported by MICINN Projects Nos.\ FIS2010-21781-C02-02 and BFU2010-21847-C02-02; Generalitat de Catalunya grant Nos.\ 2009SGR838 and 2009SGR1055; the Ram\'on y Cajal program of the Spanish Ministry of Science; ICREA  Academia prize 2010, funded by the Generalitat de Catalunya.

{\it Note added.}--During the final writing of this paper, we became aware of a recent preprint where the SIR model on interconnected networks is studied~\cite{dickinson:2012}. In that work, the authors find a mixed phase where one network propagates the epidemic while the other does not. Neither our analytic nor simulation results indicate the existence of such mixing phase in our case.

\end{document}